\documentclass[]{article}
\usepackage{cite}
\usepackage{array}
\usepackage{amsmath}
\usepackage{amsbsy}
\usepackage{amssymb}
\usepackage{textcomp}
\usepackage{gensymb}
\usepackage{siunitx}
\usepackage{multirow}
\usepackage[english]{babel}
\usepackage[utf8]{inputenc}
\usepackage{booktabs} 
\usepackage{longtable}
\usepackage{graphicx}
\usepackage{subcaption}
\usepackage{cleveref}
\usepackage{fixltx2e}
\usepackage{stfloats}
\usepackage{url}
\usepackage{cite}
\usepackage{authblk}

\begin{document}
\title{A soft robotic tongue to develop solutions to manage swallowing disorders}

\author[1]{Marco~Marconati}
\author[1]{Silvia~Pani}
\author[2]{Jan~Engmann}
\author[2]{Adam~Burbidge}
\author[3]{Marco~Ramaioli}
\affil[1]{University of Surrey}
\affil[2]{Nestl\'e Research Center}
\affil[3]{INRAE}

\maketitle
\begin{abstract}
The development of novel texture adaptations for the management of swallowing disorders could be accelerated if reliable \textit{in vitro} tests were made available. This study addresses some of the limitations of swallowing \textit{in vitro} models, by introducing a soft robotic actuator inspired by the tongue. 
The wettability of the soft-robotic actuator was engineered to achieve physiologically relevant contact angles to allow comparing dry and lubricated conditions. The actuator design and the control algorithm are designed to offer flexibility in the swallowing patterns to consider different scenarios, including poor lingual coordination. \textit{In vitro} swallowing tests with shear thinning liquids were performed as a proof of concept and showed physiologically relevant oral transit time, bolus velocity and palatal pressure. Both the bolus rheology and the coordination of the peristaltic contractions influenced the temporal evolution of the bolus velocity and the bolus transit time. This novel soft robotic tongue, its integration in an \textit{in vitro} oral cavity and its flexible control can therefore contribute to developing better solutions to manage swallowing disorders.
\end{abstract}




\section{Introduction}
Swallowing is a key physiological function accomplished by more than 25 pairs of muscles \cite{Mittal2011}. Neurological diseases, muscular frailty (sarcopenia) and other conditions can lead to increasing difficulties with food manipulation and swallowing \cite{Leonard2013,Cichero2006}. 

Several clinical studies have been devoted to the assessment and management of swallowing disorders (dysphagia), identifying texture adaptation and bolus rheology as key factors to increase swallowing safety. Thickened beverages and pureed food represent therefore the most widely used compensatory technique for the daily management of oropharyngeal dysphagia.

Alongside clinical research, a number of \textit{in vitro} models have also been proposed to replicate some of the specific features of swallowing. These models proved useful to gain a more mechanistic understanding of the oral, pharyngeal and esophageal phases of swallowing \cite{Qazi2017c,Hayoun2015a,Marconati2017b,Protip2018}. However, the extent to which \textit{in vitro} models can be used as a reliable complement to sensory and clinical tests has been questioned by some researchers, in particular with regards to 1) the strong simplification of the geometry of the oral cavity and of the tongue, 2) the use of rigid materials 3) the often neglected role of salivary lubrication and 4) the choice of imposing bolus displacements rather than imposing a peristaltic pressure to the bolus \cite{Marconati2019review}.

This motivates the development of more realistic \textit{in vitro} models of swallowing that can also be used to screen novel, easy-to-swallow, product formulations.

Recent technological advances in soft robotics led to the development of actuators able to compress, elongate, and bend in multiple directions, upon pneumatic inflation of embedded air cavities \cite{Hughes2016,Cho2009,Kang2013,Lekakou2015,Cianchetti2018}, using materials with elastic moduli between 0.01 to 1000 MPa \cite{Rus2015}. 
These characteristics inspired us to use soft robotics to simulate the tongue, guiding the bolus towards the pharynx during the oral phase of swallowing. 

Pneumatic-actuated manipulators are controlled by the inflation/deflation pressure of embedded air chambers that define the kinematics \cite{Siefert2018}. 

Despite the relative ease of manufacturing of soft manipulators, the use of materials such as PDMS, rubber, or other elastomers, can present challenges related to the long-term performance under fatigue \cite{Hughes2016}.

Only few applications of soft-robotics to swallowing have so far been proposed. Chen and Dirven first introduced a soft pneumatic model of the esophagus to replicate primary esophageal peristalsis \cite{Dirven2014a,Chen2014}. The \textit{in vitro} model of Fujiso \textit{et al.} was instead developed to study the bolus flow through the phaynx  \cite{Protip2018}. This prototype, however, imposed displacements on the bolus and does not consider the oral phase of swallowing and the role of the tongue.
More recently, a soft prototype to mimic specific tongue gestures, has also been presented \cite{Lu2017}, but this model is not specifically designed to simulate the swallowing function.
Redfearn and Hanson \cite{Redfearn2018a} proposed a two dimensional soft mechanical model of tongue-palate compression with a single degree of freedom, in which a deformable tongue is displaced vertically against a rigid palate, squeezing a bolus.

The present study extends the application of soft-robotics to investigate the bolus flow in the oral phase of swallowing. This \textit{in vitro} model is designed around a soft inflatable actuator inspired by the tongue. This actuator is able to apply a peristaltic pressure wave to a liquid bolus within a three dimensional \textit{in vitro} oral cavity. In order to approximate the effect of salivary lubrication in vivo, the surface properties of the \textit{in vitro} model were engineered to achieve a similar wettability as the human tongue. The pressure pattern applied to the bolus can be controlled to simulate different swallowing scenarios.
The experimental setup is semi-transparent to allow video-recordings of the bolus movement and is equipped with probes to measurements the palatal pressures and an ultrasound probe to measure the bolus velocity at the exit of the \textit{in vitro} oral cavity.
\begin{figure*}
\centering
\includegraphics[width=1.0\textwidth]{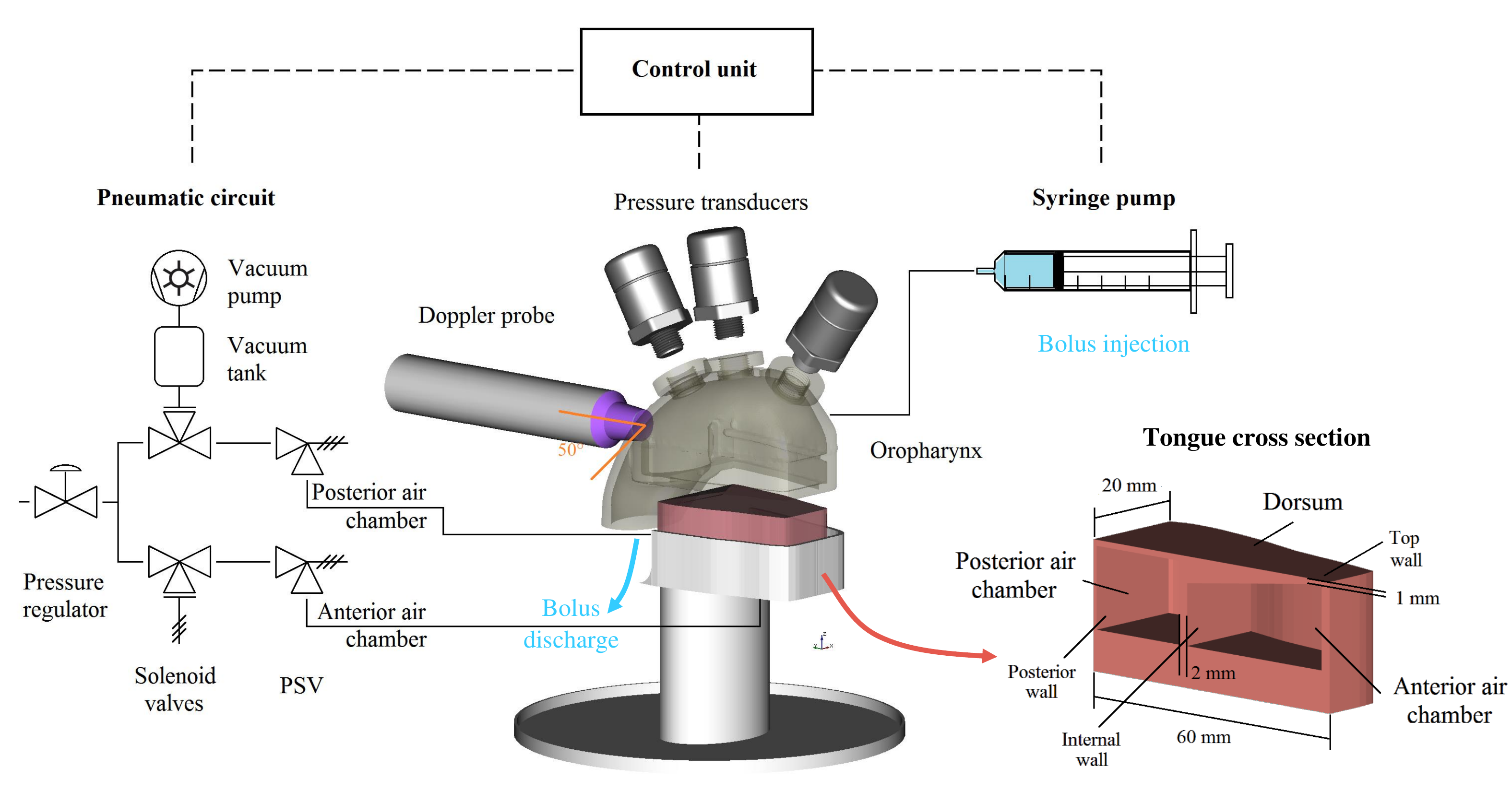}
\caption{Exploded view of the \textit{in vitro} setup. A cross section of the soft inflatable tongue is pictured to the right.}
\label{Setup}
\end{figure*}

\section{Design and fabrication}

A schematic illustration of the soft robotic tongue and of the whole experimental setup used to simulate the oral phase of swallowing is reported in Fig.~\ref{Setup}. The main components are described in the following paragraphs.

\subsection{Palate and pharynx}
A primary driver for the development of the \textit{in vitro} model came from the need to consider a more realistic geometry for the oral cavity. This requirement led to adopt a fully three dimensional model of the oropharynx.
The design of this model was based on an available 3D mesh, originally developed by Xi \textit{et al.} to characterize the fluid dynamics and deposition of aerosols \cite{Xi2007}.
Only small modifications were necessary to include the functionality of the tongue, that was not considered in the study by Xi \textit{et al.}.
Based on the image analysis of available clinical data for healthy individuals, the original 3D mesh was cut along the sagittal plane at 3 cm from the upper edge of the palate. This section was extruded and offset to offer adequate room in the longitudinal (60 mm) and transversal (40 mm) directions for housing the soft actuator.
The geometry of the palate proposed by Xi \textit{et al.} \cite{Xi2007} was only altered by adding three 1/8'' holes, located at 6, 32, 55 mm from the tongue tip along the sagittal direction. These holes were used to securely fit the pressure transducers used to record the dynamic evolution of palatal pressure during the \textit{in vitro} swallowing tests.
The oropharynx was 3D printed using an Objet 260 Connex 3 printer (Stratasys Inc., Eden Prairie, MN) with a vertical resolution of 200 $\mu$m. A partially transparent material (VeroClear) was used to observe the bolus movement during swallowing.

\subsection{Soft robotic tongue}

This study proposes a novel, simple, soft robotic tongue design that does not try to replicate the full spectrum of the tongue movements, but focuses solely on tongue peristalsis.

Ultrasound imaging of the tongue and its dynamics during swallowing \cite{Mowlavi2016} inspired the design of an actuator with two chambers that can be inflated independently. 
The length (60 mm) and width (40 mm) of this soft robotic tongue is comparable to the characteristic active length of the human tongue during swallowing \cite{Tasko2002}.

The soft robotic tongue was mold-casted with a commercially available silicone rubber (Smooth-On Eco-flex 00-30). This elastomer, already adopted by Chen and Dirven in the manufacturing of their pneumatically operated esophagus \cite{Dirven2014a,Chen2014}, is also widely used in the fabrication of soft manipulators.
The mechanical properties of Eco-flex 00-30 were characterized using dog-bone shaped specimens half the size prescribed by ASTM D412 \cite{ASTMInternational2006}. Tensile tests were performed using a texture analyser TA.XT Plus (Stable Micro Systems, Godalming, UK) equipped with a 5 kg load cell. These tests were run at a constant rate of 1 mm/s with a total set travel distance of 150 mm. Video recordings during the tests were acquired and post-processed by image analysis to calculate the strain at the mid-section of the specimens. 
Mechanical hysteresis was characterized with three consecutive runs. The measurements were also repeated after 1 and 3 weeks to evaluate the material ageing.

The wettability of the soft robotic tongue was engineered to imitate the lubrication induced by the salivary film \textit{in vivo}. Previous studies show that the surface of the human tongue is hydrophobic and weakly polar \cite{Ranc2006}. To decrease the strong hydrophobicity of silicone rubber, an anionic surfactant, sorbitan mono-oleate (Span\textsuperscript{\textregistered} 80, CAS:1338-43-8, supplied from Sigma Aldrich), was incorporated by shear mixing in the polymeric matrix before completion of the curing process. 
The surfactant concentration was adjusted so that the wettability matched the available \textit{in vivo} data. To this extent, the contact angle of DI water was measured using a drop shape analyser (model DSA30B from KRÜSS GmbH, Hamburg, Germany) on rectangular mold-casted samples of silicone rubber with different concentrations of dispersed surfactant (0.5-2 \% w/w). Measurements were repeated 6 times on different points of the samples, averaging the left and the right hand side contact angles. 
As with the tensile tests, measurements of contact angle were repeated after respectively 1 and 3 weeks to verify whether hydrophobic recovery was observed.

Two air cavities are contained within the soft actuator (Fig.~\ref{Setup}). Their inflation and deflation enables the soft actuator to replicate the two key lingual functions of 1) bolus containment prior to swallowing and 2) bolus propulsion during the oral phase of swallowing.


The anterior air chamber was designed with a characteristic trapezoidal cross section (Fig.~\ref{Setup}) to ensure an higher vertical deformation of the front tip, inducing an initial anterior contact between the tongue and the palate that progressively extends posteriorly, producing the desired backward propulsion of the liquid bolus.

The design of the internal chambers was refined through finite element (FE) simulations using COMSOL Multiphysics 3.5a (COMSOL Inc., Stockholm, Sweden), considering the mechanical properties obtained during the tensile tests and a contact pair between the inflatable tongue and the rigid palate. The inflation of the internal cavities is simulated at steady state, imposing different pressures inside the air cavities. 

\subsection{Control of the swallowing pattern}

The soft robotic tongue control has been designed to provide with the flexibility needed to simulate different swallowing scenarios, including poor lingual coordination.

An algorithm was developed to 1) automatically deliver the bolus to the oral cavity prior to swallowing, 2) control the actuation of the two chambers of the soft robotic tongue and 3) to log the data during swallowing and implemented with an Arduino Uno board in the Arduino IDE v1.8.7 programming environment (Arduino AG, Chiasso, Switzerland). The pneumatic system that controls the deformation of the soft tongue is regulated via two solenoid valves G356A002SBA8 (ASCO Numatics Sirai Srl, Bussero, Italy) and a rotary vane vacuum pump. Two adjustable relief valves, set at different pressures, allow reaching two different target inflation pressures for the anterior and posterior chambers, whilst using a single compressed air supply at 50 kPa (Fig.~\ref{Setup}).

\begin{figure*}
\centering
\includegraphics[width=1\textwidth]{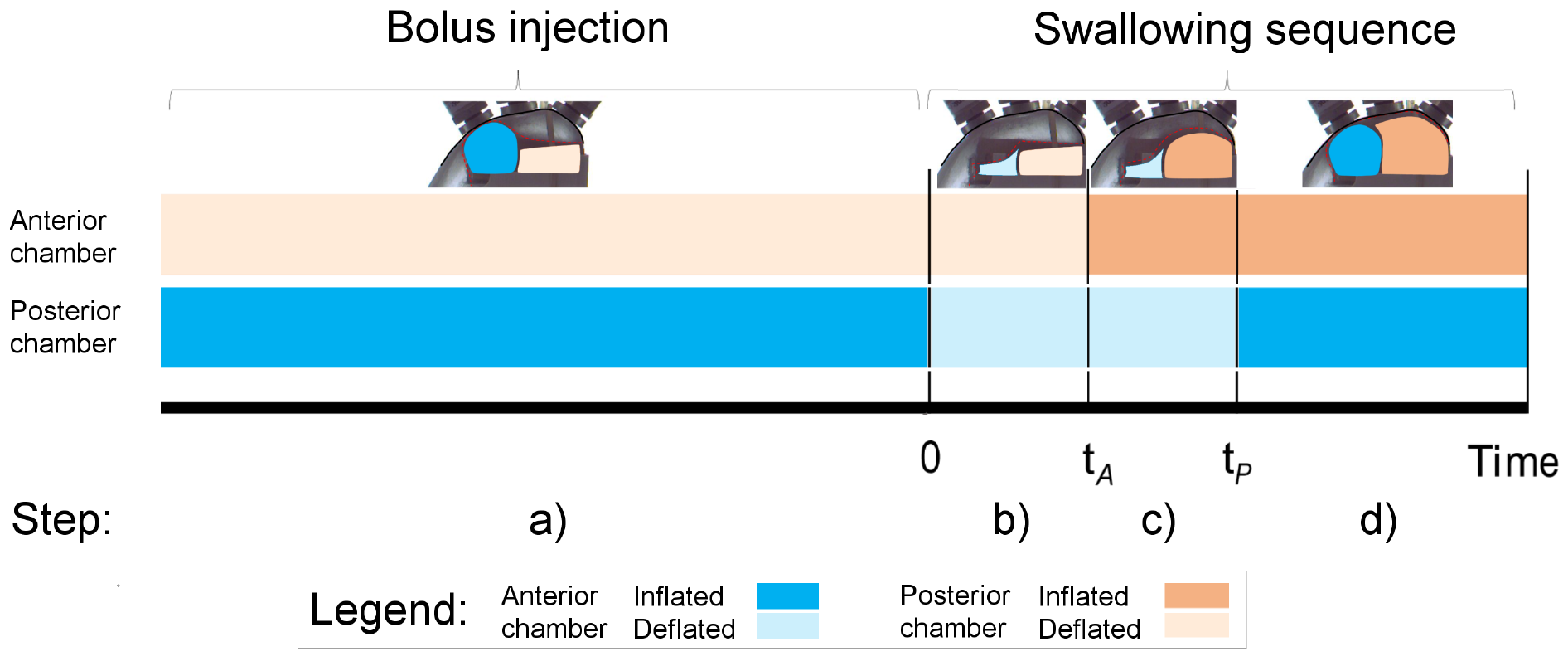}
\caption{Chronogram of the swallowing sequence of the soft robotic tongue. Lateral images are also reported to illustrate the key steps: a) bolus injection, b) deflation of the posterior air chamber, c) inflation of the anterior air chamber and d) termination of the sequence to clear the bolus residues. Colored contours illustrate the shape of the actuator along the sagittal (or medial) plane. The pneumatic actuation can be used to simulate the partial loss of lingual coordination by changing the time delay for inflation of the anterior and posterior chambers (t$_A$ and t$_P$).}
\label{Sequence2}
\end{figure*}

The actuation sequence starts with the inflation of the posterior air chamber (set at an internal pressure of 10 kPa). This seals the oral cavity posteriorly to ensure the containment of the bolus, similarly to the \textit{in vivo} bolus preparation prior to swallowing  (Fig.~\ref{Sequence2} step a) ). 
While the palate is sealed posteriorly, a syringe pump (model NE-501 from ProSense BV, Oosterhout, The Netherlands), automatically injects a controlled volume of the liquid bolus, through a 4 mm circular orifice located on top of the hard palate. Upon completion of bolus loading, suction generated by a rotary vane pump is used to generate a rapid deflation of the posterior air chamber thus leaving the bolus posteriorly unconstrained  (Fig.~\ref{Sequence2} step b) ). This sudden deflation is followed by the inflation of the anterior air chamber (set at 25 kPa) which squeezes the bolus toward the back of the oral cavity (Fig.~\ref{Sequence2} step c) ). The combined deflation/inflation of the two air chambers propels peristaltically the bolus towards the pharynx. 
A rapid re-inflation of the posterior air chamber, with the anterior one still fully inflated, is used to clear the bolus tail from the palate. After, both chambers are deflated and the system returns to its initial undeformed configuration. 

The multiple degrees of freedom of this model allow controlling independently the inflation pressures of the two air chambers and the vacuum pressure to deflate the posterior air chamber. Moreover, the control algorithm used to drive the actuator gives flexibility in the definition of different swallowing patterns, in order to ascertain the importance of tongue coordination. The duration of the swallowing sequence can be adjusted by varying the delay between the deflation of the posterior chamber and the inflation of the anterior chamber (t$_{A}$). The delay before the second inflation of the posterior cavity (t$_P$) can also be modified.

The overall duration of the \textit{in vivo} swallowing sequence reported by Tasko \textit{et al.} \cite{Tasko2002} was used to select the characteristic time-delays used for this preliminary study.


\subsection{Measured variables}
The velocity and the pressure within the liquid bolii were measured during the \textit{in vitro} swallowing tests, to enable a comparison against available \textit{in vivo} data.

\begin{figure}[ht]
\centering

\includegraphics[width=0.8\textwidth]{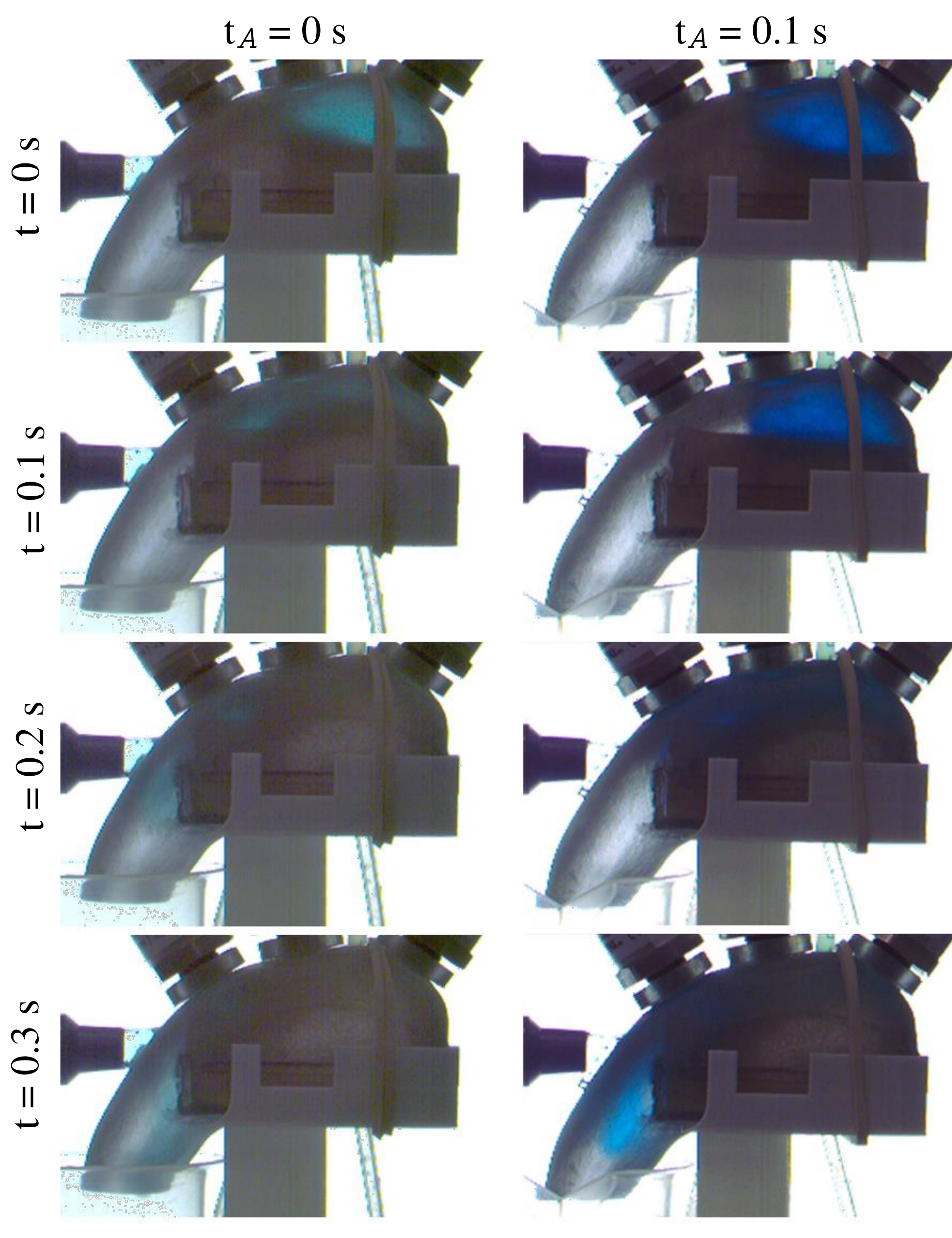}
\caption{Effect of the delay of inflation of the anterior air chamber (t$_A$=0 and 0.1 s) on the flow of an aqueous solution of 4.59\% w/w TUC.}
\label{ScreenshotsNew2}
\end{figure}

A Basler RGB camera (model ac1920-155 $\mu$m, Basler, Germany) was used to record the bolus transit at 100 frames per second. The higher time resolution, compared to the standard 24 fps of most videofluoroscopy swallowing studies, enables to capture in greater detail the bolus motion and offers a reliable mean of evaluating the characteristic oral transit time, defined as the time required for the bolus front to exit the lower part of the \textit{in vitro} oropharynx (Fig. \ref{ScreenshotsNew2}).

Doppler ultrasound is used to obtain a semi-quantitative information about the bolus velocity distribution. 
A linear Doppler probe (Dopplex IIMD2, Huntleigh Healthcare Ltd, Cardiff, Wales, UK) was installed on the posterior part of the \textit{in vitro} palate at an angle of 50\degree with respect to the direction of flow (Fig.~\ref{Setup} and \ref{ScreenshotsNew2}). 
This probe, commonly used to measure blood flow, works by emitting a continuous 5 MHz waveform and returns its echo. The amplitude of this signal is linearly proportional to the Doppler frequency shift ($\Delta f$) expressed by $\Delta f=2~f_0~{\bar{v}}/{c}~\cos \varphi$, where $f_0$ is the frequency of the piezoelectric emitter (5 MHz), $\bar{v}$ the average velocity in the field of view of the probe and $\varphi$ the angle of insonation. The speed of sound in the liquid media, $c$, was measured using an ultrasonic echo-scope (GAMPT-scan AG, Germany). The probe was calibrated using a rig imposing different liquid flow rates and considering both steady and transient flows.

Aside from the bolus kinematics, three piezoresistive pressure sensors (model PX2AG2XX002BAAAX from Honeywell, MN, USA) were housed in the three holes of the rigid palate and used to determine the liquid pressures involved in the oral transit of the bolus. The transducers selected for the study provide a linear response proportional to the absolute pressure measured and were calibrated in the range of the reported physiological pressures during the oral phase of swallowing (0 to 20 kPa).

Finally, the mass of bolus ejected from the \textit{in vitro} model was also measured after every swallow to calculate the mass of residues. The discharged bolus mass was normalized with respect to the injected bolus mass.


\subsection{Bolus properties and \textit{in vitro} swallowing conditions}
Two aqueous solutions of a commercial thickener commonly used in the management of dysphagia (Resource\textsuperscript{\textregistered} ThickenUp\texttrademark Clear, by Nestl\'e Health Science, in the following TUC) were used for this study. The thinnest solution was prepared adding 100 mL of deionized water to 1.2 g of TUC (1.19\% w/w TUC) and was categorized Level 2 under the standardization framework for dysphagia diet management established by the IDDSI \cite{Cichero2016}. A significantly thicker solution -IDDSI Level 4- was prepared by addition of 4.8 g of TUC to 100 mL of DI water (4.59\% w/w TUC).
In both cases the liquid solutions were mixed with a magnetic stirrer for 1 h at ambient temperature and used within 24h from preparation. 

The steady shear viscosity of these solutions was measured in triplicates at room temperature (22 \degree C), using a controlled stress rheometer Paar Physica UDS 200 (Anton Paar GmbH, Graz, Austria) and a cone and plate geometry (d=75 mm $\alpha$=2$\degree$).
The two solutions showed a shear thinning behavior over the range of shear rates considered (0.1 to 500 reciprocal seconds) and their viscosity could be captured well by a Power-Law
model $\eta_a$=1.82 $\dot{\gamma}^{-0.72}$ and 6.15 $\dot{\gamma}^{-0.74}$ Pa.s.

A set of swallowing tests was performed to demonstrate the performance of this soft robotic prototype when 1) varying the bolus rheology and 2) altering the coordination of the tongue peristalsis.
A 10 mL bolus volume was consistently used for all the \textit{in vitro} swallowing tests. This amount of liquid is frequently used in videofluoroscopy swallowing studies \cite{Leonard2013} and is comparable to a natural sip-size in adults \cite{Alsanei2015}.
For each liquid, the swallowing sequence was repeated five times. The first swallow started by filling the dry oral cavity with 10 mL of the desired solution, while the following swallows were run without rinsing the model (repeated swallows).


To approximate the effect of a salivary lubrication, the model was thoroughly washed from bolus residues flushing DI water through the bolus loading hole and performing three dry swallows to clear the excess water. A normal swallowing sequence was then performed (water pre-wetted swallow).

Finally, some tests were also performed to assess the sensitivity of the device with respect to variations in the actuation sequence to simulate poor tongue coordination. Specifically, the delay for the inflation of the anterior chamber was changed with 0.1 s step increments from t$_A$ = -0.2 to t$_A$ = +0.2 s. Negative values of t$_A$ are illustrative of poor tongue coordination with an anticipated inflation of the anterior air chamber when the posterior air chamber is still inflated obstructing the bolus flow toward the pharynx.
The inflation pressure of the anterior and posterior air chambers were set respectively to 25 kPa and 10 kPa for all the experiments, while the deflation pressure of the posterior air chamber was set to -20 kPa. These values led to physiologically sound pressures in the liquid, as discussed later in the text.

\section{Results}

\subsection{Mechanical and interfacial properties of the soft pneumatic tongue}

The tensile tests showed that the elastomer used to mold cast the tongue has an elongation at breakage higher than 700\% and a characteristic hyper-elastic behavior. 
%
The force displacement curve deviates significantly from a Hookean behavior for strains above 20\%. A mechanical hysteresis was observed upon removal of the applied load after the first stretch-relaxation cycle, as reported in literature \cite{Case2015}, while the following deformation cycles did not show a hysteresis.

The addition of surfactant modified both the mechanical and the surface properties of the silicone rubber (Table \ref{ContactAngle}). While at  low surfactant concentrations (0.5\% w/w), the tensile response was virtually unaffected, a significant reduction of the Young's modulus was observed for concentrations of surfactant above 1\% w/w, leading to a 3-fold decrease at the highest concentration.
The contact angle was observed to decrease sharply with low surfactant concentrations (Table \ref{ContactAngle}). A 0.5 \% w/w concentration of Span\textsuperscript{\textregistered} 80 was found sufficient to achieve the typical values of contact angles of saliva-lubricated tongue reported in the literature (approx. 50\degree) \cite{Ranc2006}. Increasing the concentration of surfactant did not show a significant further reduction in the contact angle measurement, but rather led to a softening. These observations led to select a Span\textsuperscript{\textregistered} 80 concentration of 0.5\% w/w for the final construction of the soft robotic tongue. Only a small hydrophobic recovery has been observed over three weeks, with the contact angle of the robotic tongue increasing from an average of 46\degree to 57 \degree.
The soft robotic tongues were replaced after three weeks.

\begin{table}[]
\footnotesize
\centering
\caption{Physical properties of different materials considered during the development of the soft robotic tongue. Young's modulus (E) and contact angle (CA) of Eco-Flex 00-30 with different concentrations of sorbitan mono-oleate (Span\textsuperscript{\textregistered} 80).}
\label{ContactAngle}
\begin{tabular}{cccc}
\toprule
Span\textsuperscript{\textregistered} 80 (\% w/w) & E (kPa) & CA (DEG) & \\ \midrule
0 & 77.0 (8.6) & 103.7 (4.8) & \\
0.5 & 74.7 (3.9) & 45.7 (6.1) & \\
1 & 51.6 (3.6) & 48.9 (5.5) & \\
2 & 25.3 (1.6) & 46.0 (6.6) & \\ \bottomrule
\end{tabular}
\end{table}

\subsection{Numerical simulations}

Two dimensional simulations of the tongue's sagittal cross section were performed during the design phase, to evaluate qualitatively the deformation induced by the internal air pressure. These simulations led to identifying four key geometric parameters affecting significantly the tongue deformation: 1) the thickness of the superior wall of the anterior chamber, 2) the thickness of the wall dividing the two chambers, 3) the thickness of the posterior wall and 4) the length of the posterior chamber.

Reducing the thickness of the top wall of the anterior air chamber close to the the tongue tip produces a stronger deformation, with the anterior part of the tongue reaching a contact with the palate well before the dorsum. This motion, highly desirable for bolus propulsion, can however generate a kink in proximity of the posterior air chamber. Such geometrical discontinuity can be minimized by decreasing the thickness of the wall that separates the two air chambers or by increasing the length of the posterior air chamber. However, alterations of the latter are not desirable and would influence the amount of liquid volume that can be held in mouth before triggering the swallowing sequence. 

\begin{figure}
\centering

\includegraphics[width=0.8\textwidth]{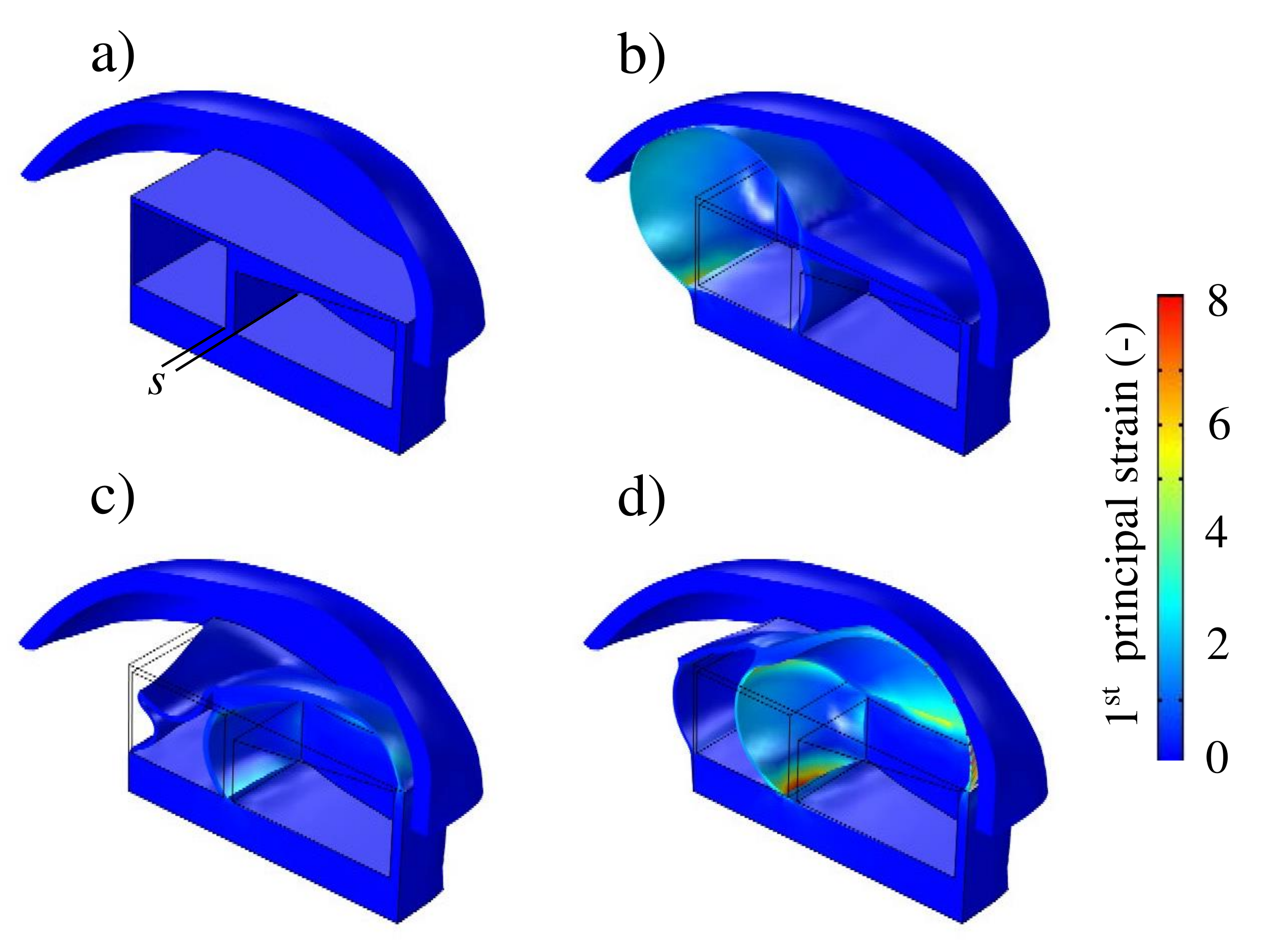}
\caption{Three dimensional FE simulation of tongue actuation. a) undeformed configuration, b) inflation of the posterior part of the tongue, c) start of the swallowing sequence, d) inflation of the anterior air chamber.}
\label{Comsol2}
\end{figure}

A second key design variable was identified with the thickness of the wall that separates the two chambers of the actuator (labeled $s$ in Fig.~\ref{Comsol2}). In light of the high strains that this region has to withstand, a sensitivity analysis was performed considering a full 3D geometry to complement 2D simulations. When $s$ was increased above 2.5 mm an evident discontinuity was generated at the joint between the two air chambers. Conversely, lower $s$, led to a higher local strains that could compromise the deformation pattern required for efficient bolus propulsion. As a result of these factors, a 2 mm thickness was selected for the wall separating the two chambers.

The thickness of the posterior wall of the soft robotic tongue is also an important dimension to ensure a smooth bolus transit, especially when the posterior chamber gets deflated under vacuum. Increasing the wall thickness limits vertical buckling and generates a saddle on the dorsum of the tongue (Fig.~\ref{Comsol2} c) and in some cases also an undesirable dead volume that is not realistic with respect to the tongue physiology. Conversely, reducing the wall thickness can lead to a complete buckling of the posterior part of the tongue thus reducing the longitudinal curvature of the saddle when the posterior air chamber is deflated (Fig.~\ref{Comsol2}). The collapse of the posterior wall under buckling leaves also a wider gap between the tongue and the palate, thus enabling a smoother transit of the bolus.

\subsection{\textit{In vitro} swallowing tests}

\subsubsection{Oral transit time and residues}

The \textit{in vitro} model was capable of simulating the swallowing of both liquids into the pharynx, without any uncontrolled leakage and respecting the physiological duration of the oral phase of swallowing ($<$0.5 s). Transit times of approximately 0.3 s were measured for both TUC solutions, despite the different concentration and viscosity. 
This observation is in agreement with the inertial dynamics experienced by highly shear thinning boli, as opposed to thick Newtonian fluids, already discussed by Mowlavi \textit{et al.} \cite{Mowlavi2016}.
The transit time of the products was consistent across repeats, confirming the repeatability of the actuation sequence. 

The amount of oral residues was calculated by measuring the discharged bolus mass through the \textit{in vitro} oropharynx. The transiting bolus mass decreases for  increasing bolus consistency. The IDDSI Level 4 solution (4.59 \% w/w TUC) left significantly more residues after the first swallow (i.e. dry mouth) than the thinner 1.19 \% w/w TUC . Specifically, measurements indicate that 87$\pm$3 \% of the injected bolus mass for 1.19 \% w/w TUC was effectively transported and ejected during \textit{in vitro} swallowing tests, leaving approximately  1.3 g of oral-residues coating the dorsum of the tongue and the oral cavity. Conversely, at the highest concentration, 73$\pm$5 \% of the injected bolus mass was discharged and the mass of residues for the dry-mouth condition increased to 2.5 g.

The introduction of water as a model lubricant to did not lead to any measurable variation in the transit time. However, the discharged bolus mass increased (+7\%) using TUC 1.19 \% w/w thus suggesting a reduction of oral-residues in case of a water lubricated swallow. In the case of 4.59 \% w/w TUC such reduction was less significant. These results call for further tests, considering lubricating films that can reproduce more accurately the peculiar rheological and tribological properties of human saliva. 


\subsubsection{Doppler velocimetry}

The magnitude of the intra-bolus velocity during swallowing was obtained from Doppler ultrasound spectra and this provides an insight on the bolus transport efficiency.
The velocity profiles typically presented two main peaks within the duration of the swallowing sequence (t=0-0.6 s). The first peak corresponds to the bolus front entering the field of view of the probe. The signal then progressively decreased as the bolus transited between the posterior palate and the root of the tongue. A second, narrower, peak was observed when the posterior chamber re-inflated, clearing the remaining liquid.
The velocity measurements obtained from the ultrasound signals appeared realistic with respect to \textit{in vivo} measurements ($<$0.5 m/s) \cite{Kumagai2009} and the two TUC solutions showed comparable velocities, in agreement with \cite{Mowlavi2016}.
The effect of water lubrication on the measured Doppler signal was not strong and could not be clearly distinguished from the experimental error.


\subsubsection{Palatal pressures}

Several clinical studies investigated the maximum isometric palatal pressure and attempted to characterize the dynamic variation of the palatal pressure during swallowing. A general agreement was found in the maximum pressures that can consciously be applied by the tongue against the palate. However, less consensus exist on the pressures applied during swallowing, probably due to the different transducers used, different contact conditions and the interpersonal variability. Redfearn and Hanson, compared the maximum palatal pressures indicated by ten studies and reported a range of pressures spanning between 1 to 43.5 kPa, with a median of 13.6 kPa \cite{Redfearn2018a}.

The \textit{in vitro} model presented in this article is driven by imposing pressures. This approach limits the maximum applicable pressure to the hard palate, conversely to imposing strains, or displacements, which might lead to non realistic palatal pressures.

\begin{figure*}[t!]
\centering
\begin{subfigure}[t]{0.45\textwidth}
\caption{\label{PressureTUC12}}
\includegraphics[width=1\textwidth]{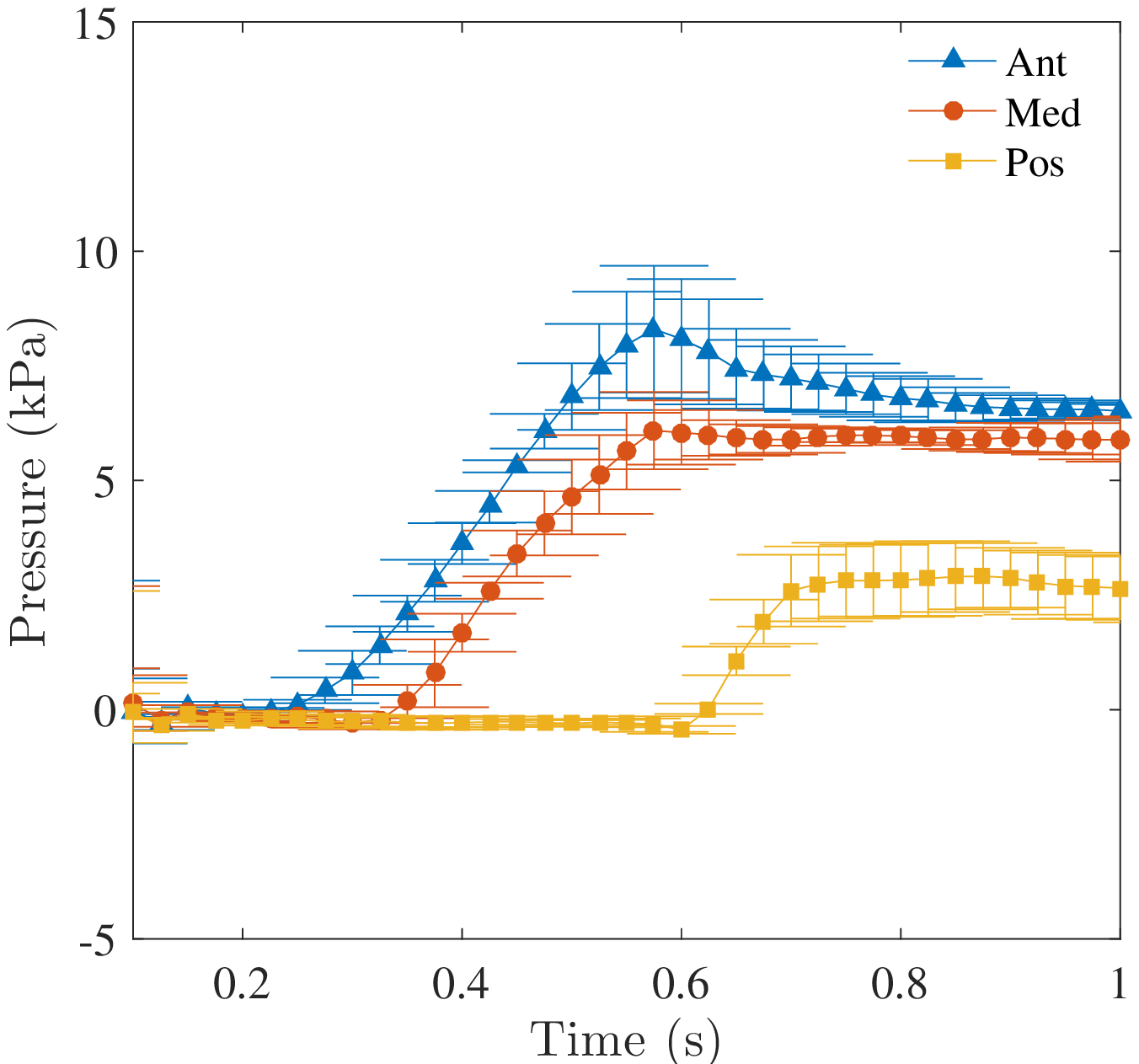}
\end{subfigure}%
~
\begin{subfigure}[t]{0.45\textwidth}
\caption{\label{PressureTUC48}}
\includegraphics[width=1\textwidth]{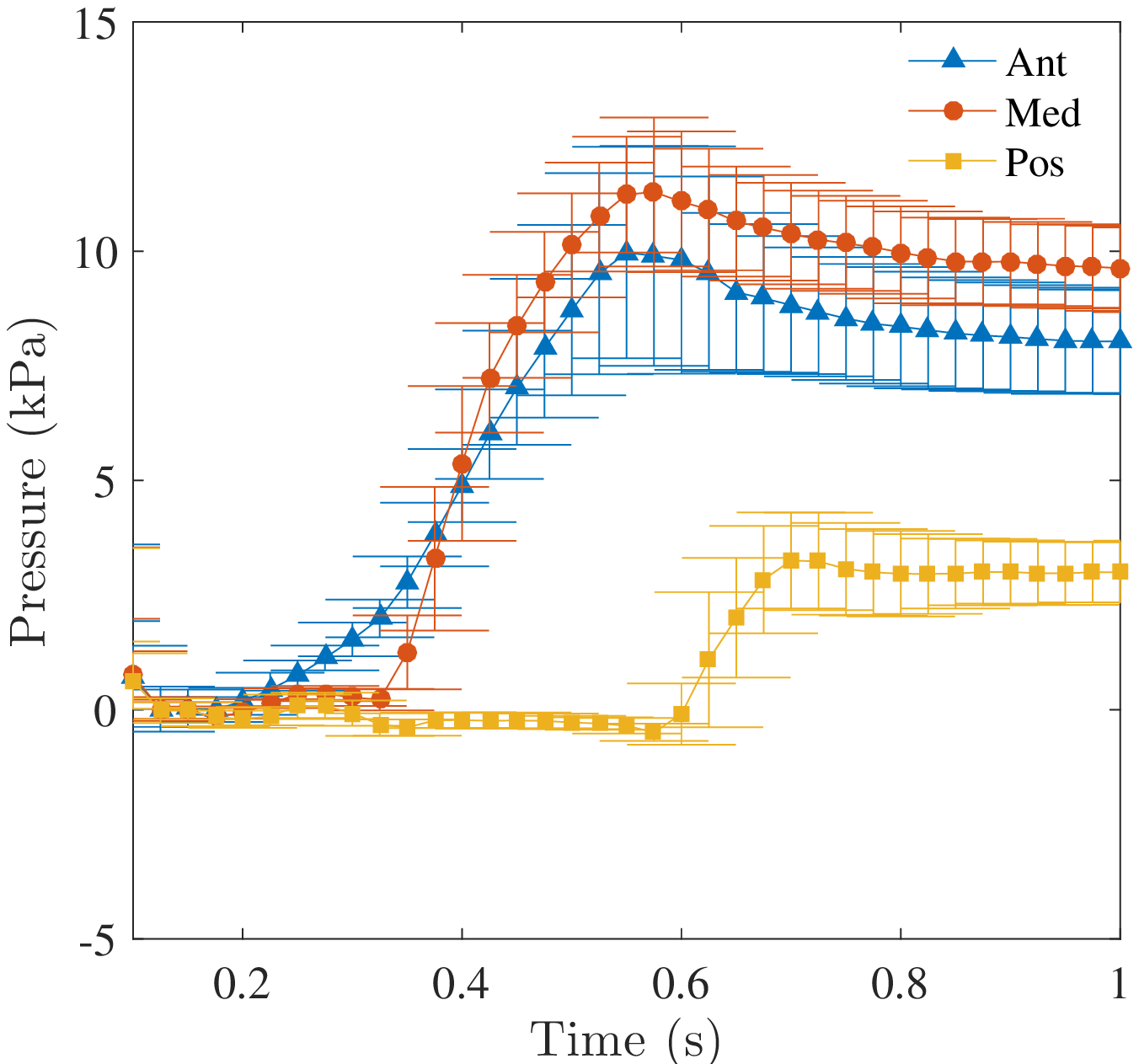}
\end{subfigure}

\caption{Pressures measured at the anterior, median and posterior palate from \textit{in vitro} swallowing tests with the aqueous solutions of TUC: a) 1.19 \% w/w TUC, b) 4.59 \% w/w TUC. In the swallowing sequence t$_A$ was set to 0.1 s and t$_P$=0.5 s.}
\label{PressureTUC}
\end{figure*}

The pressure profiles measured \textit{in vitro} (Fig.~\ref{PressureTUC}) compare favourably with the \textit{in vivo} results obtained with healthy patients in the study by 
Hashimoto et al. \cite{Hashimoto2014}. In particular, the anterior pressure grows in both cases more rapidly than the pressure at the median and posterior positions. Furthermore the maximum peak reached by the anterior pressure is higher than reached at the median position and higher than the peak at the posterior, both \textit{in vitro} and \textit{in vivo}.

The \textit{in vitro} results suggest that palatal pressures increase with bolus consistency. The pressure distribution increases at the front of the palate after approximately 0.2 s from the deflation of the posterior chamber and this delay can be controlled by changing t$_A$.
The onset of the pressure wave on the anterior palate is promptly followed by a sharper increase in the pressure measured at the median position (Fig.~\ref{PressureTUC}). This part of the pressure profiles can be put in relation with the low values of intra-bolus pressures measured by Redfearn and Hanson in their quasi bi-dimensional model of tongue-palatal squeezing  \cite{Redfearn2018a}.

The pressure gradient in the anterior-posterior direction decreases sharply when the two profiles intersect (Fig.~\ref{PressureTUC48}). Video recordings of the experiment show that this event occurs after the bulk of the bolus has left the \textit{in vitro} oral cavity.
The further increase in the measured pressures after the bolus transit is the result of the remaining film of liquid being squeezed against the palate and probably be related to the effort required to clear the oral residues, rather than to the actual bolus transport. 
Water lubrication had a minor impact on the pressure profiles and these results are not included for brevity.


\subsection{Simulating a lack of tongue coordination}

Clinical studies showed that the kinematics and the dynamics of tongue motion is subject to a high degree of interpersonal variability, both in terms of duration and amplitude. Unsurprisingly, that variability is further amplified by neurogenic dysphagia \cite{Leonard2013}.

To demonstrate the potential of this novel \textit{in vitro} prototype to provide insight on the effect of poor tongue coordination, tests with different values of the parameter t$_A$ were performed. These experiment focused on the thickest solution of TUC (4.59\% w/w) due to the clinical importance of thick shear thinning liquids in the management of oropharyngeal dysphagia.
Remarkably, \textit{in vitro} tests show no premature spillage of the bolus even for a 0.2 s delayed inflation of the anterior chamber. This result confirms the benefits of thick and shears thinning products for the compensation of poor coordination, thus promoting a better bolus control prior and during swallowing.

\begin{figure}
\centering
\includegraphics[width=0.8\textwidth]{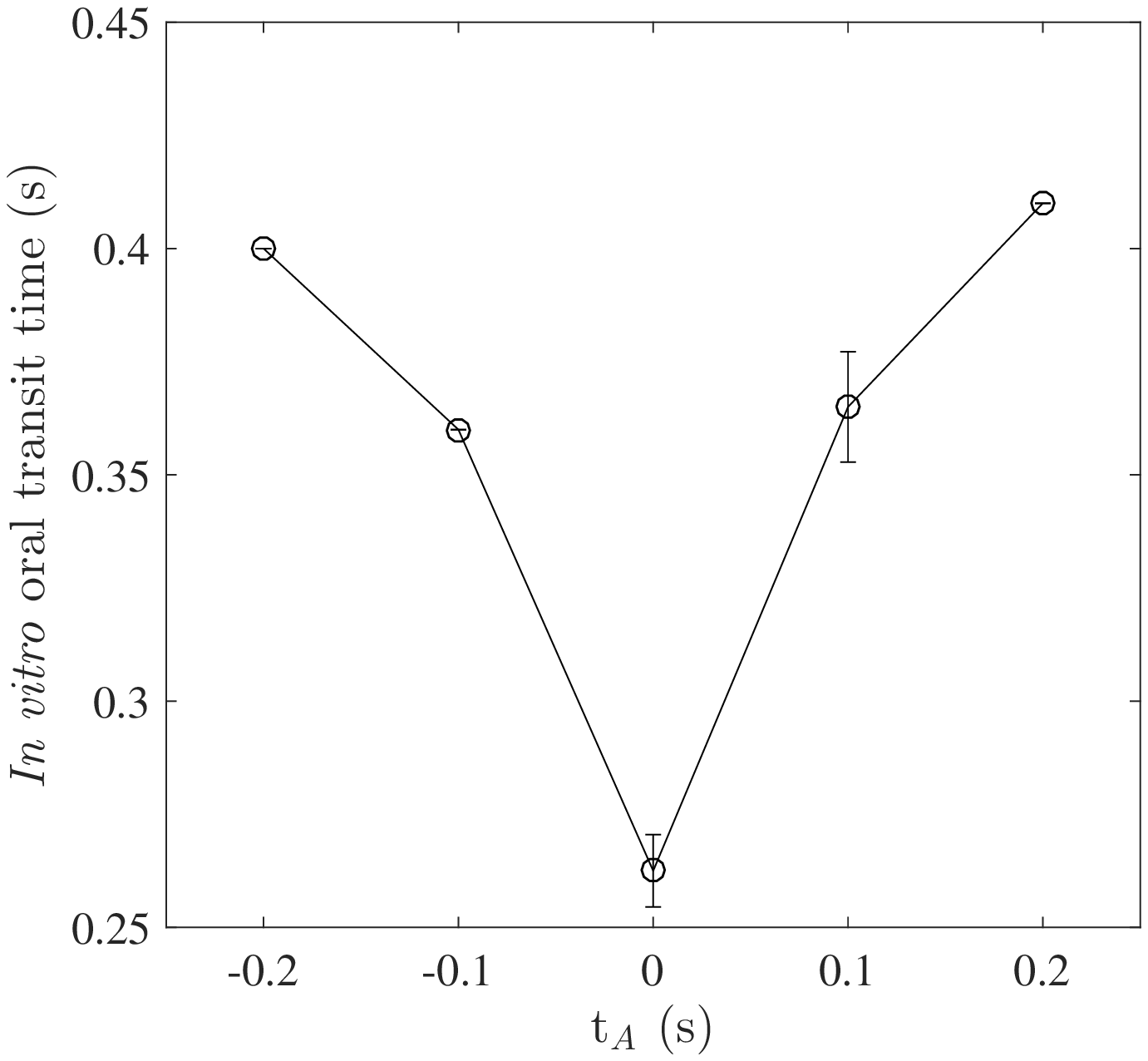}
\caption{Oral transit times from video recordings of the experiment for different time delays of bolus propulsion (t$_A$ from -0.2 to 0.2 s). The thickest TUC solution was used as a model liquid bolus.}
\label{Timingtable}
\end{figure}

\begin{figure}
\centering

\includegraphics[width=0.8\textwidth]{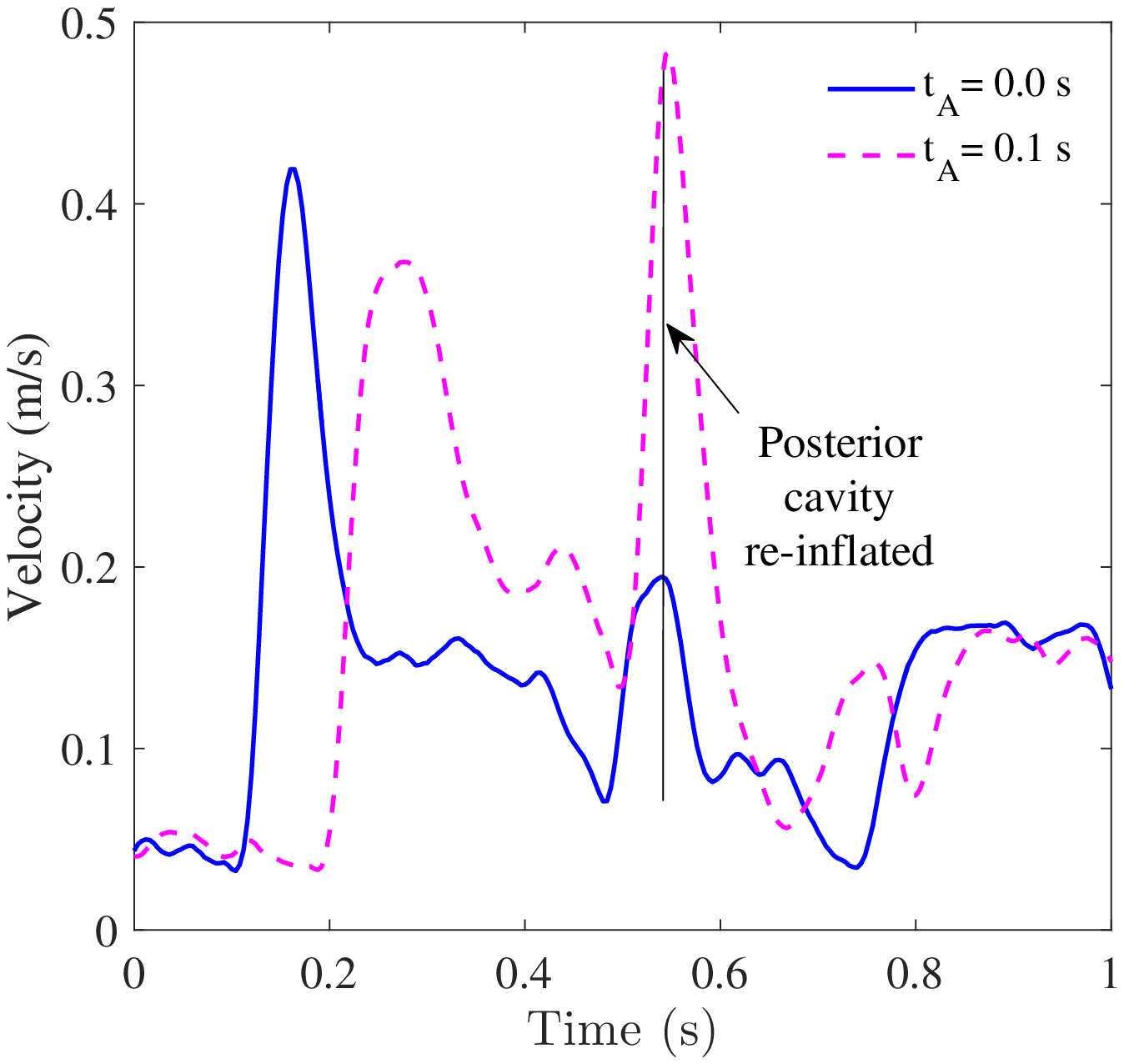}
\caption{Doppler velocity profiles from \textit{in vitro} swallowing experiments for different time delays of bolus propulsion (t$_A$=0 and 0.1 s, solid and dotted lines respectively). The thickest TUC solution was used as a model liquid bolus.}
\label{DopplerTiming}
\end{figure}

The experimental results showed also a non monotonous dependence of oral transit time on t$_A$. The shortest transit time (0.27 s) was measured in case of a perfect coordination, when the bolus propulsion delay was zero. Longer transit times were instead measured for poor coordination, whether this is caused by a delayed or an anticipated contraction of the anterior part of tongue.

A slower dynamics is expected as t$_A$ increases. The increase in the measured oral transit time is however lower than the increase in t$_A$ imposed via the control. When t$_A$ was increased from 0.1 to 0.2 s, the oral transit time increased only from an average of 0.37 to 0.41 s (Fig.~\ref{Timingtable}).

Conversely, when the inflation of the anterior chamber was triggered before the deflation of the posterior chamber (negative t$_A$), the duration of oral transit still increased as a consequence of the deceased efficiency of peristaltic transport. In this scenario the lack of coordination results in the bolus being confined in a partially closed cavity, thus inducing a higher viscous dissipation.

The Doppler profiles provided a semi-quantitative indication of the efficiency of the bolus transport. A positive shift in the onset of the Doppler signal was observed for a delayed inflation of the anterior chamber (as shown in Fig.~\ref{DopplerTiming}, comparing the values for t$_A$=0 s and 0.1 s).
%
Both Doppler signals showed a characteristic second peak before 0.6 s caused by the re-inflation of the posterior chamber, clearing the remaining bolus residues from the back of the oral cavity.
Interestingly, the amplitude of this second peak is significantly smaller when the delay was set to 0 s, suggesting that a perfect synchronization of the anterior inflation and posterior deflation lowers the liquid residues left on the posterior part of the palate when the posterior contraction occurs (Fig.~\ref{DopplerTiming}). 
Similar results were obtained when t$_A$ was set to -0.1 s, although the velocity was lower and the distribution wider.

Despite these differences in the bolus velocity, the discharged bolus mass (and therefore the amount of oral residues) was not strongly affected by the delay of inflation of the anterior chamber. Other parameters might affect more strongly the bolus clearance from the oral cavity, such as the inflation pressure of the anterior and posterior chambers and their effect should be investigated in future studies.

\section{Conclusion}
This study proposes a novel approach to study the oral phase of swallowing introducing a soft-robotic actuator inspired by the human tongue. 
The swallowing dynamics of a wide range of bolus consistencies can be studied in this model and preliminary tests have confirmed that physiologically sound liquid pressures can be applied. \textit{In vitro} swallowing tests, at imposed stresses, showed an increase in palatal pressures when increasing bolus consistency and higher pressures applied at the front of the tongue.

This soft robotic tongue offers unique capabilities to study the effect of oral lubrication and related disorders (xerostomia). The wetting properties have been engineered to simulate the role of oral lubrication. The absence of lubricant was shown to increase slightly the residues with thin boli, but had a negligible effect on transit time. This motivates further research to consider more realistic models of lubricating liquids to mimic the role of the salivary film.

The effect of a poor tongue coordination on swallowing can be investigated thanks to the actuator degrees of freedom and control. Such capability can be used to search for novel rheological profiles that could mitigate this effect and result in a more robust bolus and safer flow.

\section*{Acknowledgment}
This study was sponsored by Nestl\'e Research.


\bibliographystyle{IEEEtran}

\bibliography{Collection_R2}

\end{document}